\documentstyle[12pt]{article}
\begin{document}
\font\rm=cmr12
\font\tenrm=cmr10
\font\tensl=cmsl10
\font\tenbf=cmbx10
Preprint N.   INFN-ISS   93/5    September 93
\begin{center}
{}~~\\[5cm]
\tenrm
{\tenbf  INCLUSIVE QUASIELASTIC AND DEEP INELASTIC
 SCATTERING OF POLARIZED ELECTRONS BY POLARIZED $^{3}$He}\\
[0.8cm]
\rm{C. Ciofi degli Atti and S. Scopetta}\\
{\tensl Department of Physics, University of
Perugia, and INFN, Sezione di Perugia,\\
Via A. Pascoli, I-06100 Perugia, Italy}\\[0.3cm]
\rm{E. Pace}\\
{\tensl Dipartimento di Fisica, Universit\`a di Roma "Tor Vergata",
and INFN, \\
Sezione Tor Vergata, Via E. Carnevale, I-00173 Roma, Italy}
\\[0.3cm]
\rm{G. Salm\`e} \footnotetext {Presented by G. Salm\`e}
\\
{\tensl INFN, Sezione Sanit\`a, Viale Regina Elena 299, I-00161
Roma, Italy}
\\[.8cm]
\end{center}
\centerline{\underline {Abstract}}
\medskip
A comprehensive treatement of the
theoretical approach for describing  nuclear effects in
inclusive scattering of polarized electrons by polarized $^3$He is
presented. \bigskip

\centerline{\underline {1. Introduction}}

\medskip

\indent  The advent of new
experimental facilities, allowing  systematic measurements with
polarized $^{3}$He and polarized electrons beams, are substantially increasing
the amount  of information
on the electromagnetic properties of the neutron, in a wide range of the
kinematical variables. As is well known, polarized $^{3}$He
represents a good candidate as an effective neutron target [1].
The effects of  nuclear structure, however, have to be carefully
investigated in order to reliably extract information on the
neutron from the data  both in the quasielastic (qe) [2,3]  and in
the deep inelastic [4] region. In what follows, we will illustrate
how  the proton contribution affects the extraction
of the neutron elastic form factors  [5-7] and  spin structure
functions [8].  \bigskip

\pagestyle{plain}
\centerline{\underline{ 2. The  polarized inclusive cross section}} \medskip

\indent The inclusive cross section describing the scattering of a
longitudinally
polarized lepton of helicity $h~=~\pm1$ by a polarized hadron of spin J = 1/2,
is given in one photon exchange approximation by [1a]
\begin{eqnarray}
\frac{d^2\sigma(h)}{d\Omega_2 d\nu}\equiv
\sigma_2\left(\nu,Q^2,\vec{S}_A,h\right) = \frac{4 \alpha^2}{Q^4}\;
\frac{\epsilon_2}{\epsilon_1}\;m^2\;
L^{\mu\nu}W_{\mu\nu}= \frac{4 \alpha^2}{Q^4}\;
\frac{\epsilon_2}{\epsilon_1}\;m^2\;
\left[L^{\mu\nu}_{s}W_{\mu\nu}^{s}+L^{\mu\nu}_{a}
W_{\mu\nu}^{a}\right]
\label{eq1}
\medskip
\end{eqnarray}
where $L^{\mu\nu}_{s(a)}$ and $W_{\mu\nu}^{s(a)}$ are the symmetric ($s$)
(antisymmetric ($a$)) leptonic and hadronic tensors, respectively.
The antisymmetric hadronic tensor is given
by
\begin{eqnarray}
 W_{\mu\nu}^{a}
&=&\;i\epsilon_{\mu\nu\rho\sigma}q^\rho\;V^\sigma
 \label{eq3.2}
\medskip \end{eqnarray}
where $V^\sigma$ is a pseudovector   that can be
expressed as follows
\begin{equation}
V^\sigma\equiv~S^\sigma_A~\frac{G^A_1}{M_A} +
(P_A\cdot q\;S^\sigma_A - S_A\cdot q\;P^\sigma_A)
{}~\frac{G^A_2}{M^3_A}\label{eq3.3}
\medskip
\end{equation}
In the above equations, the index A
 denotes the number of nucleons  composing
the target; $G^A_1$ and $G^A_2$ are the
polarized structure functions;  $k^\mu_{1(2)} \equiv
(\epsilon_{1(2)},\vec{k}_{1(2)})$ and $P_A^\mu  \equiv (M_A,0)$
are electron and target four-momenta;  $~q^\mu \equiv
(\nu,\vec{q})$ is the four-momentum transfer, $Q^2 = -q^2$;
$~g_{\mu\nu}$  is the symmetric metric tensor,
$\epsilon_{\mu\nu\rho\sigma}$ the fully antisymmetric tensor and
$~S_A^\mu$  the polarization four-vector (in the rest frame
$S_A^\mu \equiv (0,\vec{S}_A)$).

\indent The antisymmetric hadronic tensor, $W^{a}_{\mu\nu}$,  is constrained
 to the very general expression (\ref{eq3.2}) by
invariance principles (Lorentz, gauge,	parity and time reversal
invariance); moreover the antisymmetric tensor $\epsilon_{\mu\nu\alpha\beta}$
in
Eq.(\ref{eq3.2}) cancels out  any contribution to $W^{a}_{\mu\nu}$ arising from
possible terms  proportional to $q^\mu$ ($\epsilon_{\mu\nu\alpha\beta}~
q^\alpha~q^\beta~=~0$) in $V^\sigma$. This fact has to be carefully taken into
account once a model is adopted for obtaining the polarized structure functions
[6].

\indent The polarized
structure functions  $G_1^A$ and   $G_2^A$ have to be obtained by expressing
them  in terms of the components of  $W^{a}_{\mu\nu}$. Thus,
  assuming in the rest frame
of the target the z-axis
along the momentum transfer  ($\hat{q}~\equiv~\hat{u}_z$), and using
Eq.(\ref{eq3.3}),  one has (cf. Refs.[6,7])
\begin{eqnarray}
{G_1^A\over M_A}~=~-i \left({Q^2\over |\vec{q}|^3}~{W^{a}_{02}\over
S_{Ax}}~+~{\nu \over |\vec{q}|^2} {W^{a}_{12}\over S_{Az}}\right)
{}~~~~~~~
{G_2^A\over M_A^2}~=~-i{1\over |\vec{q}|^2}~\left({\nu \over
|\vec{q}|}{W^{a}_{02}\over S_{Ax}}~-~ {W^{a}_{12}\over S_{Az}}\right)
\label{5.2}\medskip
\end{eqnarray}

\noindent  It should be pointed out
that in Ref.[1a] $G_{1(2)}^A$ have been obtained by another procedure,
namely using  the components of the pseudovector ${V}^\sigma$,
Eq.(\ref{eq3.3}); in this case one has
\begin{eqnarray}
{G_1^A\over M_A}~=~-{(V\cdot q)\over{|\vec{q}|S_{Az}}}
{}~~~~~~~~~~
{G_2^A\over M_A^2}~=~{V_0\over{|\vec{q}|S_{Az}}}
\label{5.4}
\medskip
\end{eqnarray}
Given the form (\ref{eq3.3}) for $V^\sigma$, Eq.(\ref{5.2})
are totally equivalent to Eq.(\ref{5.4}). However, such an
equivalence will{\em{ not hold }}if a term proportional to $q^\mu$
is explicitely added to the r.h.s. of Eq.(\ref{eq3.3}), since
Eq.(\ref{5.2})  will be unaffected by the added term, whereas Eq.
(\ref{5.4}) will be; therefore  $G_1^A$ and  $G_2^A$ obtained
from  Eq.(\ref{5.4})  will  not be correct in this case [6].

\indent After contracting
the two tensors in Eq.(\ref{eq1}) one has
\begin{eqnarray}
\frac{d^2\hbox{$\large\sigma$}(h)}{d\Omega_2 d\nu}~=~
\hbox{$\large\Sigma$}\;+\;h\;\hbox{$\large\Delta$}
\label{eq14}
\end{eqnarray}
where $\large\Sigma$ and $\large\Delta$ describe the unpolarized
and the polarized scattering, respectively. The polarized term is
\begin{eqnarray}
\hbox{$\large\Delta$}~=~
\hbox{$\large\sigma$}_{Mott}~2~tan^2\frac{\theta_{e}}{2}
\left[\frac{G^A_1(Q^2,\nu)}{M_A}~(\vec{k_1}+\vec{k_2})~+~2~
\frac{G^A_2(Q^2,\nu)}{M_A^2}~(\epsilon_1
 \vec{k_2}-\epsilon_2 \vec{k_1})\right]\cdot\vec{S_A}
\label{eq14.2}
\medskip
\end{eqnarray}

Experimentally one measures the  following asymmetry
\begin{eqnarray}
\rm A= \frac {{\sigma }_{2}\left({\nu ,{Q}^{2},{\vec{S}}_{A},+1}\right) -
{\sigma
}_{2}\left({\nu ,{Q}^{2},{\vec{S}}_{A},-1}\right)} {{\sigma }_{2}\left({\nu
,{Q}^{2},{\vec{S}}_{A},+1}\right) + {\sigma }_{2}\left({\nu
,{Q}^{2},{\vec{S}}_{A},-1}\right)} =
\frac{\hbox{$\large\Delta$}}{\hbox{$\large\Sigma$}}
 \label{eq18}
\end{eqnarray}

\indent In order to obtain the theoretical asymmetry,
one has to introduce some  approximations;
in particular, till now, the Plane Wave Impulse Approximation (PWIA) has been
adopted. Within such a framework  the polarized structure functions $G_1^A$ and
$G_2^A$ are given by (cf.  Refs.[6-7] for the qe case and [8-9] for the deep
inelastic one)

\begin{eqnarray}
{}~~~~\frac{G^A_1\left (Q^2,\nu\right )}{M_A}=
 \sum\nolimits\limits_{N=p,n} \int \nolimits \nolimits~dz
\int \nolimits \nolimits~dE \int \nolimits
\nolimits~d{\vec p}~ {1\over E_p~M}~\left \{
\hat{G}_{1}^{N}\!\left(z,\nu,{Q}^{2} \right) \left[M~{\hbox{\Large\it
P}}_{\parallel}^{N}\!\left( |{\vec p}|,E,\alpha\right)~+
\right. \right. \nonumber \\
\left.\left.\;-|{\vec p}|\left({\nu\over|\vec q|}
-{|{\vec p}|cos\alpha\over{M+E_p}}\right)
{}~\hbox{$\Large\cal P$}^{N}\!\left(|{\vec p}|
, E,\alpha \right)\right]-{Q^2\over|\vec q|^2}~{\hbox{$\Large\cal
L$}}^{N} \right \}~\delta \left(z+{M^2-p \cdot p\over 2M \nu} -{q \cdot p \over
M \nu}\right)
 \label{eq11}
\end{eqnarray}

\begin{eqnarray}
\frac{G^A_2\left (Q^2,\nu\right )}{M_A^2}=
  \sum\nolimits\limits_{N=p,n}
\int \nolimits \nolimits dz
\int \nolimits \nolimits dE \int \nolimits
\nolimits d{\vec p}~{1\over E_p~M}~\left \{ \left [
\hat{G}_1^{N}\!\left(z,\nu,{Q}^{2} \right)~{|{\vec p}| \over
|\vec{q}|}~ {\hbox{$\Large\cal P$}^{N}\!\left(|{\vec
p}|,E,\alpha\right)}\;+\right.\right. \nonumber \\
\left.\left.+~{\hat{G}_{2}^{N}\!\left(z,\nu,{Q}^{2}
\right)\over M} \left(E_p~{\hbox{\Large\it
P}}_{\parallel}^{N}\!\left(|{\vec p}|,E,\alpha\right)-{|{\vec
p}|^2~cos\alpha\over{M+E_p}} ~\hbox{$\Large\cal P$}^{N}\!\left(|{\vec p}| ,
E,\alpha\right)\right)\right]~-~{\nu\over|\vec
q|^2}~{\hbox{$\Large\cal
L$}}^{N} \right \}
\nonumber \\
\delta \left(z+{M^2-p \cdot p \over 2M \nu} -{q \cdot p \over
M \nu}\right )~~~~~~~~~~~~~~~~~~~~~~~~~~
\label{eq12}
\end{eqnarray}

\noindent where E is the removal energy,
$p\equiv(M_A-\sqrt{(E+M_{A}-M)^2+|\vec
p|^2},~\vec p)$,$~E_p=\sqrt{M^2+|\vec{p}|^2}$ and $\hat{G}_{1(2)}^{N}\!\left
(z,\nu, Q^{2} \right)$ are the nucleon structure functions
to be used in the
energy transfer region considered (cf. Sects. 4 and 5, below). The function
${\hbox{$\Large\cal L$}}^{N}$ is
\begin{eqnarray}
{\hbox{$\Large\cal
L$}}^{N}&=&\left[\hat{G}_1^{N}\!\left (z,\nu, Q^{2} \right)~{\cal
H}_1^N~+~|\vec q|~{\hat{G}_{2}^{N}\!\left ( z,\nu,Q^{2} \right )
\over M}~{\cal H}_2^N\right ]
\label{12.0}
\medskip
\end{eqnarray}
where ${\cal H}_1^N$ and ${\cal H}_2^N$ are proper combinations of
$~{\hbox{\Large\it P}}_{\parallel}^{N}\!\left({p},E,\alpha\right)$ and
$~{\hbox{\Large\it P}}_{\perp}^{N}\!\left({p},E,\alpha\right)$ [6],
and $~{\hbox{$\Large\cal P$}}^{N}\!\left( {p},E,\alpha\right) =
cos\alpha~{\hbox{\Large\it
P}}_{\parallel}^{N}\!\left({p},E,\alpha\right)~+~sin\alpha~{\hbox{\Large\it
P}}_{\perp}^{N}\!\left({p},E,\alpha\right)$.
 The quantities ${\hbox{\Large\it P}}_{\parallel}^{N}\!\left( {p} ,
E,\alpha\right)$ and ${\hbox{\Large\it P}}_{\perp}^{N}\!\left( {p} ,
E,\alpha\right)$, already used in a previous paper of ours [5], are related to
the elements of the 2x2 matrix,
representing the spin dependent spectral function of a nucleon inside a
nucleus with  polarization $\vec{S}_A$. The elements of this matrix
 are
 \begin{eqnarray} P_{\sigma,
\sigma',\cal{M}}^{N} ({\vec{p},E})=\sum\nolimits\limits_{{f}_{A-1}}
{}~_{N}\langle{\vec{p},\sigma;\psi
}_{A-1}^{f} |{\psi }_{J\cal{M}}\rangle~ \langle{\psi
}_{J\cal{M}}|{\psi }_{A-1}^{f};\vec{p},\sigma '\rangle _{N}~
\delta (E-{E}_{A-1}^{f}+{E}_{A})
\label{eq9}
\medskip
\end{eqnarray}
where $|{\psi
}_{J\cal{M}}\rangle$ is the ground state of the target nucleus polarized along
$\vec{S}_A$, $|{\psi }_{A-1}^{f}\rangle$ is an eigenstate of the (A-1) nucleon
system, $|\vec{p},\sigma\rangle_N$ is the plane wave for the nucleon $N\equiv
p(n)$.

\bigskip

\centerline{\underline {4. The asymmetry in the quasielastic region}}\medskip

\indent As  is well known, the experiments in the qe region [2,3] are aimed at
investigating the neutron elastic form factor. In this
kinematical region,   the nucleon structure functions
$\hat{G}_{1(2)}^{N}\!\left (z,\nu, Q^{2} \right)$ are related to the Sachs
electromagnetic form factors $G^{N}_E$ and $G^{N}_M$ as follows
\begin{eqnarray}
\hat{G}_1^{N}\!\left (z,\nu, Q^{2} \right)&
=&- {1 \over \nu} \delta\left ( z - {Q^2 \over 2M \nu} \right)~{G_M^{N}\over
2}~{(G_E^{N}+\tau~G_M^{N})\over (1+\tau)}
 \label{7.1} \\
\hat{G}_2^{N}\!\left (z,\nu, Q^{2} \right)& =& {1 \over \nu} \delta\left ( z
- {Q^2 \over 2M \nu} \right)~{G_M^{N}\over
4}~{(G_M^{N}~-~G_E^{N})\over (1+\tau)}
\label{7.2}
\medskip
\end{eqnarray}
 with $\tau=Q^2/(4M^2)$.

By substituting Eqs.(\ref{7.1}) and (\ref{7.2}) in Eqs.(\ref{eq11}) and
(\ref{eq12}) one obtains the expressions of $G_{1(2)}^A$ given in Ref.[6].
Present experimental results aim at measuring the quantities $R^A_{T'}$ and
$R^A_{TL'}$, given by
 \begin{eqnarray}
R^A_{T'}(Q^2,\nu)&=& -2~ \left(\frac{G^A_1(Q^2,\nu)}{M_A}~ \nu~ - ~Q^2
\frac{G^A_2(Q^2,\nu)}{M_A^2}\right)
{}~=~i~2~{W^{a}_{12}\over S_{Az}}
 \label{eq17a}
\end{eqnarray}
\begin{eqnarray}
R^A_{TL'}(Q^2,\nu)&=&2~\sqrt{2}~|\vec{q}|~\left(\frac{G^A_1(Q^2,\nu)}{M_A}~
+~\nu
\frac{G^A_2(Q^2,\nu)}{M_A^2}\right)
{}~=~-i~2\sqrt{2}~{W^{a}_{02}\over S_{Ax}}
\label{eq17b}
\end{eqnarray}
The interest in these quantities is due to the fact
that $R^A_{T'}$ and $R^A_{TL'}$,  at the top of the qe peak, are
proportional to  $(G^n_M)^2$ and $G^n_E~G^n_M$, respectively, {\em{provided
the
proton contribution can be disregarded}} [2,3].

\indent In Fig. 1 the asymmetry, measured  by the MIT-Caltech collaboration
[2],
corresponding to $\epsilon_1 = 574~MeV$ and $\theta_e=44^o$  and averaged over
three different values of the polarization angles around $\theta^*\approx
90^o$,
$(\cos\theta^*=\vec{S}_A\cdot\hat{q}$)
 is shown together with the neutron (dotted line) and
proton (dashed line) contributions. It is worth noting that,
in these kinematical conditions, the measured
asymmetry reduces to $R_{TL'}$  only at the top of the qe peak
($A^{exp}_{qe}\propto~R_{TL'}^{exp}$). Therefore, as
previously explained, one could have access to $G^n_E~G^n_M$,
provided  the proton contribution can be disregarded; but
unfortunately, it is shown that this not the case, for the proton
contribution is relevant  at the  top of the qe peak.  A
comparison with the experimental value obtained after averaging
over the polarization angle and  over a $100~MeV$ interval for the
energy transfer yields  \begin{eqnarray}   A^{exp}_{qe}& = &
2.41~\mp1.29~\mp0.51~\%  ~~ MIT-Caltech^2   \nonumber\\  A^{th}& =
& 1.65~\% ~~~~~~ (3.74~\%)  \nonumber \medskip\end{eqnarray}

\begin{figure}
\vspace{5cm}
\special{picture dam1}
\vspace{0.5cm}
\parbox[t]{7.5cm}{{\tenrm	Fig. 1.  The asymmetry corresponding to  $\epsilon_1$
=
574~$MeV$ and $\theta_e~ = ~$44$^o$,  vs. the
energy transfer $\nu$ calculated by  Eqs. (10) and (11) (solid line) and using
the spin-dependent spectral function of Ref.[5]; the dotted (dashed) line
represents the neutron (proton) contribution. The nucleon
form factors of Ref. [10] have been used and the experimental data are from
Ref.[2]. The arrow indicates the position of the qe peak. (After
Ref.[6])}} \ $~~~~~~$ \ \parbox[t]{7.5cm}{
{\tenrm	Fig. 2. The total asymmetry  at the top of the qe
peak, vs. Q$^2$, for  $\theta_e~=~$75$^0$ and $\beta~=~$95$^o$,
using Eqs. (10) and (11). The  Galster form factors [11] have been
used. The curves in the lower part of the figure represent the
corresponding proton contributions. (After Ref.[6])}}  \end{figure}
\noindent The theoretical value in the brackets is obtained
without
averaging over the energy, i.e. at $\nu=\nu_{peak}$.

 \indent In correspondence to a different
choice of  kinematical variables, $\epsilon_1~=574~MeV$ and
$\theta_e~=~51.1^o$, only one  experimental point has been
obtained,  just for the asymmetry averaged over both three
polarization angles around $\theta^*\approx0^o$ and  a $50~ MeV$
interval for the energy transfer around the top of the qe peak,
where $A^{exp}_{qe}\propto~R_{T'}^{exp}$ [2]. The comparison reads
as follows \begin{eqnarray}
A^{exp}_{qe}&=&-3.79~\mp1.37~\mp0.67~\%~~MIT-Caltech^2 \nonumber\\
A^{th}&=&-4.30~\% ~~~~~~ (-3.43)~\%
\nonumber
\medskip\end{eqnarray}

 It should be pointed out
that our results are only slightly different from the ones
obtained in Ref.[7], where a  spin-dependent Faddeev spectral function for
$^3He$ and the nucleon form factors of Ref.[11] have been used.

\indent The results presented in Fig. 1 show that the proton contribution to
the
measured asymmetry  is
sizeable. However, as shown in Refs. [5,6], one can minimize or even  make
vanishing the proton contribution. As a matter of fact, it turns out that it is
possible to find a polarization angle $\beta=\beta_c$
($\cos\beta=\vec{S}_A\cdot\hat{k_1}$) in correspondence of which the proton
contribution at the top of the qe peak  vanishes within a wide range of values
of the incident electron energy, and even for different models of the nucleon
form factors.  Moreover, in order to investigate the sensitivity of the
asymmetry upon $G^n_E$, such a quantity has been calculated [6] in the range
$0.3\leq Q^2\leq 2~(GeV/c)^2$, at fixed values of $\beta_c~=~95^o$ and
$\theta_e~=~75^o$, using the Galster  form factors [11],
since  within such a model  $G_E^n$ can be changed independently of $G_M^n$. In
fact one has \begin{eqnarray} G_M^n~=~{\mu_n~G_E^p}~~~~~~~~~~~~~~
G_E^n~=~{-\tau~\mu_n\over(1+\eta~\tau)}~G_E^p  \label{19}
\medskip
\end{eqnarray}
where  $G_E^p = 1/(1+Q^2/B)^2$, $B~=~0.71 (GeV/c)^2$ and $\eta$ is a parameter.
The resulting asymmetry and the proton contribution are shown in Fig. 2 for
different values of  $\eta$. Therefore Fig.2
illustrates  how the total asymmetry can depend upon  $G^n_E$, having
a vanishing proton contribution.

\indent It should be stressed that the
proposed kinematics, which minimizes the proton contribution, corresponds to
the  qe peak, where the final state interaction is expected to play only a
minor
role.
\bigskip

\centerline{\underline {5. The asymmetry in the deep inelastic region}}
\medskip

Deep inelastic scattering (DIS) of longitudinally
polarized electrons off polarized $^3He$ aimed at measuring the
spin structure functions (SSF) of the neutron, $g_1^n$ and
$g_2^n$, whose knowledge provides information on the spin
distribution among the  nucleon partons and can allow a very important
test of QCD, e.g. a check of the  Bjorken Sum Rule [12]. As is well
known, only recently  $g^n_1$ has become experimentally available
from two different experiments [13-14] on polarized deuteron and
$^3$He.

\indent The longitudinal ($\beta=0$) asymmetry can be recast in the following
form, suitable for the analysis
of DIS (see, e.g., Refs.[9,14])
\begin{eqnarray}
A_{||}=
2x [1+R(x,Q^2)]{g_1^A(x,Q^2)-{Q^2 \over \nu (\epsilon_1 + \epsilon_2 \cos
\theta_e)} g_2^A(x,Q^2)\over F_2^A(x,Q^2)}  \label{as}
\medskip
\end{eqnarray}
where $x=Q^2/ 2M\nu$ is the Bjorken
variable, $g_1^A$ and $g_2^A$
are
the nuclear SSF, $F_2^A$ is the spin-independent
structure function of the target $A$, $R(x,Q^2)=\sigma_L (x,Q^2)/ \sigma_T
(x,Q^2)$. The expressions for  $g_1^A$ and $g_2^A$ [9] can  easily
be obtained from Eqs.(\ref{eq11}) and (\ref{eq12}) by using the
following replacements
\begin{eqnarray}
g_1^A~=~M\nu~{G_1^A\over M_A}~~~~~~~~~~~~~~~~~~g_2^A~=~M\nu^2~{G_2^A\over
M_A^2}~~~~~~~~~~~~~~~~~~~~~~~ \\
g^N_1(x,z,Q^2)~=~{{p\cdot q}\over M}~
\hat{G}_1^{N}\!\left(z,\nu,{Q}^{2}\right)
{}~~~~~~~~g^N_2(x,z,Q^2)~=~{{(p\cdot q)^2}\over M^3}~
\hat{G}_2^{N}\!\left(z,\nu,{Q}^{2}\right)
\label{sub}
\medskip
\end{eqnarray}
 In the Bjorken
limit  $(\nu /
| {\vec q} | \rightarrow 1$, $Q^2/| {\vec q} |^2 \rightarrow 0)$, the
asymmetry reduces to  \begin{eqnarray}
A_{||}=2x{g_1^A(x) \over F_2^A(x)} \label{asv}
\medskip
\end{eqnarray}
and  $g_1^A$ becomes a function of $g^N_1$ only; namely it reads as follows
\begin{eqnarray}
g_1^A(x)  =  \sum_N \int _x ^A dz
{1 \over z} g_1^N \left( {x \over z} \right)
G^N(z)~~, \label{fin}
\medskip
\end{eqnarray}
with the spin dependent light cone momentum distribution for the nucleon given
by
\begin{eqnarray}
G^N(z)  =  \int dE\, \int d {\vec p}
\bigg \{  {\hbox{\Large\it
P}}_{\parallel}^{N}\!\left({p},E,\alpha\right)- \left[ 1 -
{p_{||} \over E_p + M} \right] {|{\vec p}| \over M}
{\hbox{$\Large\cal P$}}^{N}\!\left( {p},E,\alpha\right) \bigg\}
\delta \left(z - {p^+ \over M} \right)~~~ \label{lux}
\medskip
\end{eqnarray}
where $p^+=p^0-p_{||}$ is the light cone momentum component.

The ${^3 \vec{He}}$ asymmetry (Eq.(\ref{as}))
and the SSF $g^3_1$, Eq.(\ref{fin}), calculated in the Bjorken limit  using
the SSF $g^N_1$ of Ref.[15], are shown in Figs. 3a and 3b, respectively.
We would like to stress the following point:
the non vanishing
proton contribution to the asymmetry shown in Fig. 3a
hinders in principle
the extraction of the neutron structure
function from the $ {^3\vec{ He}}$ asymmetry. It can be seen from Fig. 3b that
for $0.01 \leq x \leq 0.3$ the neutron contribution, $g^{3,n}_1$,
differs from the neutron structure function $g_1^n$
by a factor of about $10\%$; since this factor is generated by
nuclear effects, one might be tempted to consider it
as the theoretical error on the determination of $g_1^n$;
however, it should be recalled that
the difference
between $g_1^n$ and $g_1^{3,n}$
is in principle model dependent
through
the way nuclear effects are introduced
and the specific form of $g_1^n$ used in the convolution
formula. In order to investigate in detail such a question,  Eq.(\ref{fin}) has
been extensively analysed  in Refs.[8,9], where it has been shown
that a factorized formula for $g^3_1$ (see Ref.[16] for the qe region)
represents a reliable approximation of the Eq.(\ref{fin}) at least for
$x~\leq~0.9$. The factorized formula can be
heuristically obtained by expanding ${1 \over z} g_1^N \left( {x \over z}
\right)$  in Eq.(\ref{fin}) around $z=1$ and by disregarding the term
proportional to $\hbox{${\Large\cal P}^N$}$ in Eq.(\ref{lux}),  which gives
anyway a very small contribution, being of the order  $|{\vec p}|
/ M$. Thus one has
\begin{eqnarray}
g_1^3(x) & \approx & 2p_p g_1^p(x) + p_n g_1^n(x)\label{gmod2}
\medskip
\end{eqnarray}
where $~p_p~$ and $~p_n~$ are the effective nucleon polarizations, produced by
the $S'$ and $D$ waves in the ground state of $^3$He, and  given by
\begin{eqnarray} p_{N}& = & \int dE~\int d\vec p {\hbox{\Large\it
P}}_{\parallel}^{N}\!\left({p},E,\alpha\right)
\label{pol}
\medskip
\end{eqnarray}
Our calculations yield $~p_p=-.030~$ and $~p_n=0.88~$ in agreement with
world values  $~p_p =-0.028{\pm} 0.004~$ and $~p_n=0.86 {\pm}
0.02~$ reported in Ref.[16].  In Fig. 4, the relevant nuclear effects, due to
the effective nucleon polarizations induced by $S'$ and $D$ waves,
are illustrated through the comparison between the free neutron
structure function and the quantity  \begin{eqnarray} \tilde
g_1^n(x)={1 \over p_n} \left[ g_1^3(x)-2p_p g_1^p(x) \right]
\label{g1art} \medskip \end{eqnarray}
calculated using the convolution formula for $g_1^3(x)$. It can be seen
that the two quantities are very close to each other, differing,
because of
binding and Fermi motion effects, by at most $4\%$.
Such a small difference
is rather
independent of
the form of any well behaved $g_1^N$ [8], and therefore Eq.(\ref{gmod2}) can be
considered a workable formula for extracting $g_1^n(x)$ from the experimental
$g_1^3(x)$.

\begin{figure}
\vspace{6cm}
\special{picture dam4}
\vspace{0.5cm}
 \parbox[t]{7.5cm}{
{\tenrm	Fig. 3a. The $^3$He asymmetry [Eq.\ (\protect \ref{asv})]
calculated within the convolution
approach [Eq.\ (\protect \ref{fin})] (full line). Also shown are the
neutron (short--dashed line) and proton (long--dashed line)
contributions.
 (After Ref.[8])}} \
$~~~~~~$ \ \parbox[t]{7.5cm} {{\tenrm	Fig. 3b. The SSF $g_1^3$ of $^3$He
(full line); also shown are the neutron
(short--dashed line) and proton (long--dashed line) contributions.
The dotted curve represents the free neutron structure function
$g_1^n$. The difference between the dotted and short--dashed
lines is due to nuclear structure effects. (After Ref.[8]) }}
\end{figure}
\begin{figure}
 \parbox{6cm}{
{\tenrm	Fig. 4. The free neutron structure function $g_1^n$ (dotted line)
compared with the neutron structure function given by
Eq.\ (\protect \ref{g1art})(dashed line). The difference between
the two curves
is due to Fermi motion and binding effects. The SSF $g^N_1$ of Ref.[15] has
been used. (After Ref.[8])}} \ $~~~~~$ \ \parbox{9cm}
{\raisebox{-6cm}{\special{picture
dam5}}} \end{figure}

\bigskip

\centerline{\underline {6. Summary and conclusion}}
\medskip

\indent The analysis of the asymmetry, based on the correct expression
of $G^A_1$ and $G^A_2$ given by Eqs.(\ref{eq11}) and
(\ref{eq12}), respectively, has put in evidence : i) the relevance
of the proton both in the qe [6,7] and the DIS regions [8,9],
ii) the possibility of selecting a polarization angle, which leads at qe peak
to
an almost vanishing proton contribution for a wide range of the kinematical
variables [6], and therefore making feasible the analysis of
the sensitivity of the asymmetry to the electric neutron form factor; iii) the
reliability of the factorized formula, represented by  Eq.(\ref{gmod2}), for
extracting $g^n_1(x)$ from the experimental $g^3_1(x)$.

\indent Calculations of the final state effects are in progress.

\bigskip
\centerline{\underline{6. References}}
\medskip
\begin{itemize}
\item[1.] a) B.
Blankleider and R.M. Woloshyn, {\em{Phys. Rev.}} {\bf{C 29}}, 538  (1984);
b) R.M. Woloshin, {\em {Nucl. Phys.}} {\bf{A495}}, 749 (1985).
\item[2.] a) C. E. Jones-Woodward et
al., {\em{Phys. Rev.}} {\bf{C}} 47, 110 (1993) and references
quoted therein; b)
R.D. McKeown, this Conference. \item[3.] A. K. Thompson et
al., {\em{Phys. Rev. Lett.}} {\bf{68}}, 2901 (1992); A. M.
Bernstein, {\em{Few-Body Systems Suppl.}} {\bf{6}}, 485  (1992).
\item[5.] C. Ciofi degli Atti, E. Pace and G.
Salm\`e, {\em{Phys. Rev.}} {\bf{C 46}}, R1591  (1992).
\item[6.] C. Ciofi degli Atti, E. Pace and G.
Salm\`e, INFN-ISS 93-3, Proceedings of the VI Workshop on
"Perspectives in Nuclear Physics at Intermediate Energies", ICTP,
Trieste, May 3-7, 1993 (World Scientific, Singapore) to be
published. \item[7.] R.W. Schultze and P.U. Sauer, {\em{Phys.
Rev.}} {\bf{C 48}}, 38 (1993).
\item[8.] C. Ciofi degli Atti, E. Pace, S. Scopetta and G.
Salm\`e, {\em{Phys. Rev.}} {\bf{C 48}}, R968  (1993).
\item [9.] C. Ciofi degli Atti, S. Scopetta, E. Pace, and G. Salm\`e,
Proceedings of the VI Workshop on "Perspectives in Nuclear Physics at
Intermediate Energies", ICTP, Trieste, May 3-7, 1993 (World Scientific,
Singapore) to be published.
\item[10.] M. Gari and W. Krumpelmann, {\em Z. Phys.} {\bf A322},
689 (1985); {\em Phys. Lett.} {\bf B 173}, 10 (1986).
\item[11.] S. Galster et al., {\em Nucl. Phys.} {\bf B32}, 221
(1971). \item [12.] J. D. Bjorken, {\em{Phys. Rev.}} {\bf D 1},
1376 (1971). \item [13.] a) SMC Collaboration, B. Adera et
al.,{\em{Phys. Lett.}}  {\bf B 302}, 553 (1993); b) S. Platchkov,
this Conference. \item [14.] a) E142, P. Anthony et
al.,{\em{Phys. Rev. Lett.}} {\bf 71} 95 (1993); b) Z.E. Meziani,
this Conference.  \item [15.] A. Sch\"afer, {\em{Phys. Lett. }}
{\bf B 208}, 175 (1988).   \item [16.] J. L. Friar, B. F. Gibson,
G. L. Payne, A. M. Bernstein, and T. E. Chupp, {\em{Phys. Rev.}}
{\bf C 42}, 2310 (1990). \end{itemize}
\end{document}